\documentclass[12pt]{iopart}
\usepackage{graphicx}
\begin{document}
\title[COVID]{Optimising SARS-CoV-2 pooled testing strategies on social networks for low-resource settings}

\author{K I Mazzitello$^1$, Y Jiang$^2$,  
C M Arizmendi$^1$}

\address{$^1$ Instituto de Investigaciones Cient{\'i}ficas y Tecnol\'ogicas en Electr\'onica, Universidad Nacional de Mar del Plata, B7608 Mar del Plata, Buenos Aires, Argentina}
\address{$^2$ Department of Mathematics and Statistics, Georgia State University, Atlanta, GA 30303, USA}

\ead{kmazzite@mdp.edu.ar}


\begin{abstract}
Controlling the COVID-19 pandemic is an urgent global challenge. The rapid geographic spread of SARS-CoV-2 directly reflects the social structure. Before effective vaccines and treatments are widely available, we have to rely on alternative, non-pharmaceutical interventions, including frequency testing, contact tracing, social distancing, mask wearing, and hand-washing, as public health practises to slow down the spread of the disease. However frequent testing is the key in the absence of any alternative. We propose a network approach to determine the optimal low resources setting oriented pool testing strategies that identifies infected individuals in a small number of tests and few rounds of testing, at
low prevalence of the virus. We simulate stochastic infection curves on societies under quarantine. Allowing some social interaction is possible to keep the COVID-19 curve flat. However, similar results can be strategically obtained searching and isolating infected persons to preserve a healthier social structure. Here, we analyze which are the best strategies to contain the virus applying an algorithm that combine samples and testing them in groups \cite{Mutesa2020}. A relevant parameter to keep infection curves flat using this algorithm is the dairy frequency of testing at zones where a high infection rate is reported. On the other hand, the algorithm efficiency is low for random search of infected people.

\end{abstract}
\maketitle

\section{Introduction}
The ongoing
COVID-19 pandemic has upended the world, quickly
challenging settled assumptions and certainties. 
It is a new virus with an extraordinary efficiency in transmitting from person to person
and a rather high level of morbidity and mortality that raises with age and co-morbidities.
The non-pharmaceutical intervention of detection and isolation of infected people is a
key policy to reduce the spread of COVID-19. The aim is
to slow transmission and the growth rate of infections to avoid overburdening
healthcare systems an approach widely known as flattening
the curve. In order to identify the infected people SARS-CoV-2 tests must be performed.

However,
each diagnostic SARS-CoV-2 test costs 30-50 US dollars \cite{Medicare}. Therefore, testing many people in
a population regularly, as may be essential to flatten the curve, is beyond the reach of most low and even some mid-income countries. However, there are more efficient ways than the naive approach of testing everyone in which far fewer tests are actually needed, especially at low prevalence. It is much
more efficient to pool (or combine) samples and test them together. Group testing initially appears in a paper of Dorfman in 1943 \cite{Dorfman1943}. Other algorithms of pooling samples have been proposed recently \cite{Mutesa2020, Armendariz2020, Hanel2020, Zhu2020, Broder2020}.  
Estimating the prevalence of a virus within a community prior to widespread disease transmission may help public health officials predict when to prepare for an increase in cases.
With over sixty eight million cases in the world at time of writing this paper \cite{WHO}, this sort of screening strategy is probably not necessary at this point in the pandemic. Nevertheless, these  techniques  are likely to be valuable at the beginning of a future outbreak to track the spread of a virus across the world over time. Specially because human behaviors that perturb the human-microbial status quo may have reached
a tipping point that predicts the inevitability of an acceleration
of disease emergences \cite{Morens2020}.

On the other hand, this approach may be particularly helpful in settings where the number of infections is low and declining, and most test results are expected to be negative. For example, in a community where the infection seems to be under control and reopenings of schools and businesses are planned, pooled testing of employees and students could be an effective strategy.

Our goal in this work is to analyze the way in which different strategies of surveillance testing in a low prevalence stage, like frequency and random vs. localized search of infected people,
change
 the epidemics curve. We choose the hypercube algorithm of pool testing  \cite{Mutesa2020} in the same way as we may have chosen some other pool testing algorithm because we are not particularly interested in the efficiency of the algorithm but in the strategy of the algorithm application. A similar study to monitor whether epidemics were contained or
 became uncontrolled depending on the frequency of testing was studied with a stochastic agent-based model for SARS-CoV2 transmission \cite{Larremore2020}. 
 To investigate the effects of surveillance testing
 strategies at the population level, we used simulations to monitor whether epidemics were contained or
 became uncontrolled,
We will take a network approach to simulate the evolution of the epidemic on a society in order to study not just the frequency but also the spatial distribution of testing.
In order to study the different behavior of the epidemic when different test pooling samples are applied the social group under the epidemic is represented as families or small communities that interact with each other in a random way. We choose a sparse network to reflect the lockdown restriction.  Similar structures have been proposed in \cite{Girvan2002} for carrying out comparative
tests of different methods for community detection in complex networks.\par
The paper is laid out as follows. In section \ref{model} we define the epidemiological network model in which the connections between individuals are
modeled as static links \cite{Vespignani2012,Newman2010}, assuming the contagion as a process faster than the network evolution.
The main results are presented in section \ref{Reults}. In section \ref{without testing}, the impact of the social structure given by the network model on the spread of the disease is analyzed. This allow us to establish a frame of reference to study, in section \ref{with testing}, the optimal strategies of the pool testing based on the geometry of a hypercube, at low prevalence \cite{Mutesa2020}. Finally, in section \ref{Conclusions}, we state our conclusions.


\section{The  epidemiological model of social networks under quarantine}
\label{model}
\begin{figure*} 
\centering
\includegraphics[width=10cm]{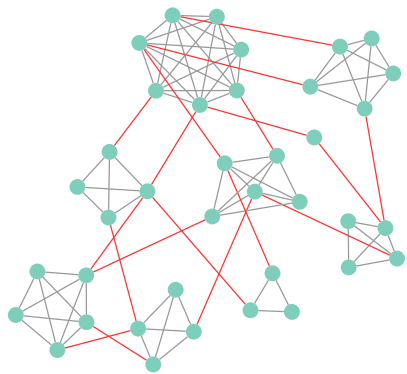}
\caption{A social network under quarantine (lockdown restriction) with few individuals ($N=38$). In this network there are nine small communities (families) consisting of $k_{int} \pm \Delta k_{int}=4 \pm 2 $ members (circles) connected to each other (gray lines) and with mean  intercommunity connections $k_{ext}=4$ per family in average (red lines). The network visualization
was created with Cytoscape \cite{Shannon2003}.
}
\label{FIG1}
\end{figure*}

In our  sparse network model, we assume that the small communities (families) are composed of a few members connected to each other and also to other families with a number of external  static links triggering the spreading epidemics (see figure  \ref{FIG1}).  The number of members of each small community or family is $k_{int} \pm \Delta k_{int}$ nodes connected in average to $k_{ext}$ nodes that belong to other small communities. The nodes of the network represent individuals that can be either susceptible, infected or recovered, 
subject to interactions with their neighbors (i.e. other individuals directly
linked to him/her by either intracommunity or intercommunity connections). As a result of these interactions, susceptible individuals can become infected and spread the disease over time before they recover or isolate using a strategy in affected areas by the virus based on the geometry of a hypercube. Starting with a number of outbreaks of the disease randomly located on the network, the model dynamics is defined by iterating a sequence of possibilities, as follows:\\

(1) an individual is selected at random;

(2a) if the individual is infected, he/she can transmit the virus to his/her neighbors with an infectious contact rate of COVID-19 pandemic $\beta$ or can recover with probability 
$1/t_{rec}$, with $t_{rec}$ the recovery time. This time is different for each infected individual, given by a Gaussian distribution around the mean value $\overline{t}_{rec}$ (see Table \ref{parameters}).

(2b) if the individual has infected neighbors, he/she can become infected with an effective contact rate $\beta$.

(3) with probability $1/t_{life}$, the selected individual can die, where $t_{life}$ is the average life expectancy. For simplicity without loss of generality, if he/she dies, an individual is born in his/her place without intercommunity connections.\\

\begin{table}[ht]
\caption{\label{parameters}Parameters and their values used in the model.}
\vspace{.3cm}
\footnotesize
\begin{tabular}{c|c|c}
\br
Parameter            & Description                      & Values\\
\mr
$\overline{\beta}$   &  Mean infectious contact rate    & $0.25\pm 0.05\;^{a}$ [1/day]\\
$\overline{t}_{rec}$ &  Mean recovery time               & $13 \pm 3.5\;^{b}$ [day]\\
$t_{life}$           &  Average life expectancy         & $75$ [year] $^{\cite{worldbank}}$\\
$k_{int}$            &  Mean number of cohabitants      & $4\pm 2\;^{d}$ \\
$k_{ext}$            &  Mean intercommunity connections & Variable \\
$p$                  &  Prevalence of the disease       & Variable \\
$N_S$                &  Number of individuals in a hypercube & Eq. (\ref{N_S})\\
$L$                  &  Size of the hypercubes          &  $3\;^{\cite{Mutesa2020}}$\\
$D$                  &  Dimension of the hypercubes     &  $L^D=N_S$   \\
$M$                  &  Maximum number of tests per day & Variable \\
$N$                  &  Maximum number of screened individuals per day & $10\times M\;^{c}$\\
$Frequency$          &  Frequency of testing and isolation &  Variable\\
                     &of infected persons per day& \\

\br
\end{tabular}\\
 $^{a}$ 95 \% confidence interval to obtain $\beta$ = 0.21-0.3 [1/day] \cite{Sanche2020}; $^{b}$ 95 \% confidence interval to obtain $t_{rec}$ = mean latent period + mean infectious period = 2.2-6 + 4-14 days = 6.2-20 days \cite{Sanche2020}. $^{c}$ Since $N$ is always greater than $M$, we consider appropriate to set $N$ one order of magnitude higher than $M$. $^{d}$ range for most countries in Latin America \cite{Household}. 
\end{table}
\normalsize

We consider that the infectious contact rate $\beta$ is constant over time but may change with different pairs of neighbors according to a Gaussian distribution around its mean value $\overline{\beta}$ (see table \ref{parameters}). Indeed, each individual experiences a different number of contacts per unit time with their neighbors, proportionally reflected in $\beta$. In other words,  $\beta$ grows with the probability of disease transmission per unit time and also with the interactions between neighbors \cite{Keeling2008,Goswami2020}.\\

The system evolves towards absorbing states with a maximum of affected individuals by the pandemic i.e., frozen configurations that are not capable of further changes. The final state, consisting of recovered and susceptible individuals that were not infected depends on the number of outbreaks of the disease and on the community structure. We are interested in studying the efficiency of a search and isolation algorithm based on the geometry of a hypercube of infected individuals on affected areas applied to the model of social networks under quarantine. The affected areas are discovered due to the rate of infected individuals report to health centres.

\section{Results and discussion}
\label{Reults}

\subsection{Epidemic spread on social networks under quarantine}
\label{without testing}

In order to set the stage for the investigation of testing effects, let us first show results concerning social network model under quarantine without any epidemic control. As mentioned above, in the absence of testing, the system reaches a total number of infected individuals that depends on the parameter values of the disease and the number of outbreaks at the beginning. These outbreaks are randomly located on the networks.\par

\begin{figure*}[h!]  
\centering
\includegraphics[width=8cm]{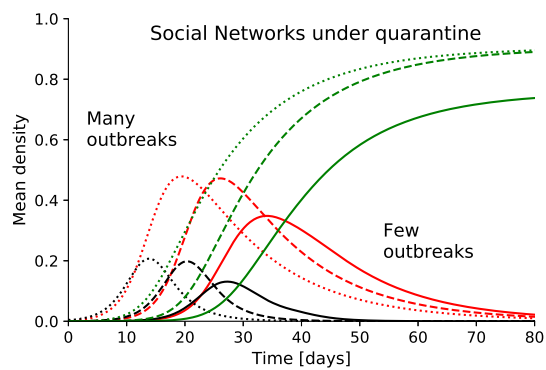} 
\vskip 0.25cm
\caption{Mean densities defined as mean numbers of recovered (green lines), active infected (red lines) and susceptible  exposed to the virus (black lines) individuals divided by the total population (100,000 inhabitants), without epidemic control, obtained from social networks under quarantine consisting of $k_{int}=4\pm 2$ cohabitants and of $k_{ext}=4$ mean intercommunity connections and different outbreaks: 1, 10 and 100 outbreaks of the virus (solid, dashed and dotted lines, respectively). The curves were averaged over 100 simulation runs. 
} 
\label{FIG2}  
\end{figure*}

\begin{figure*}  
\centering
\begin{tabular}{cc}
\includegraphics[width=8cm]{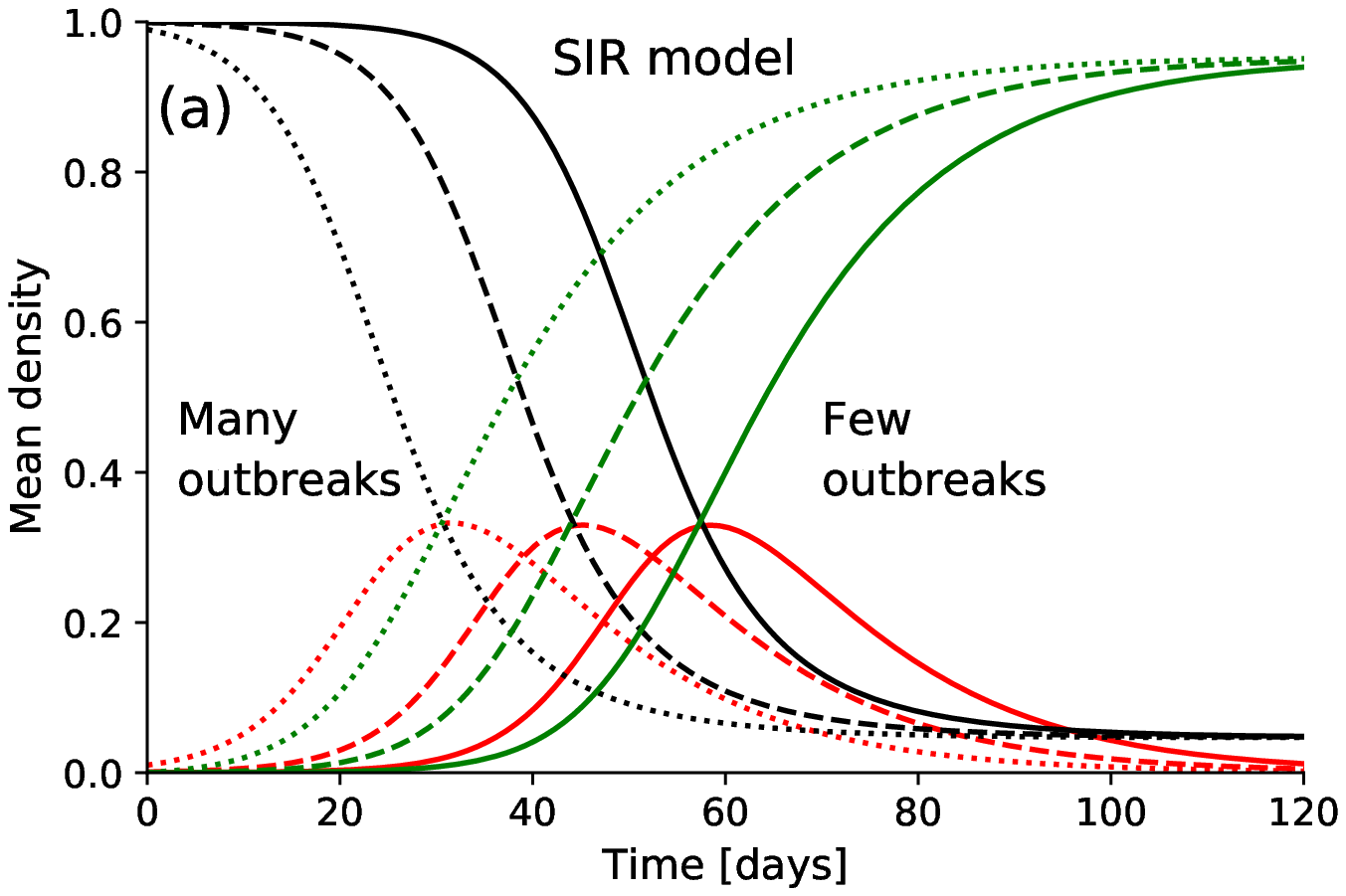}&
\includegraphics[width=8cm]{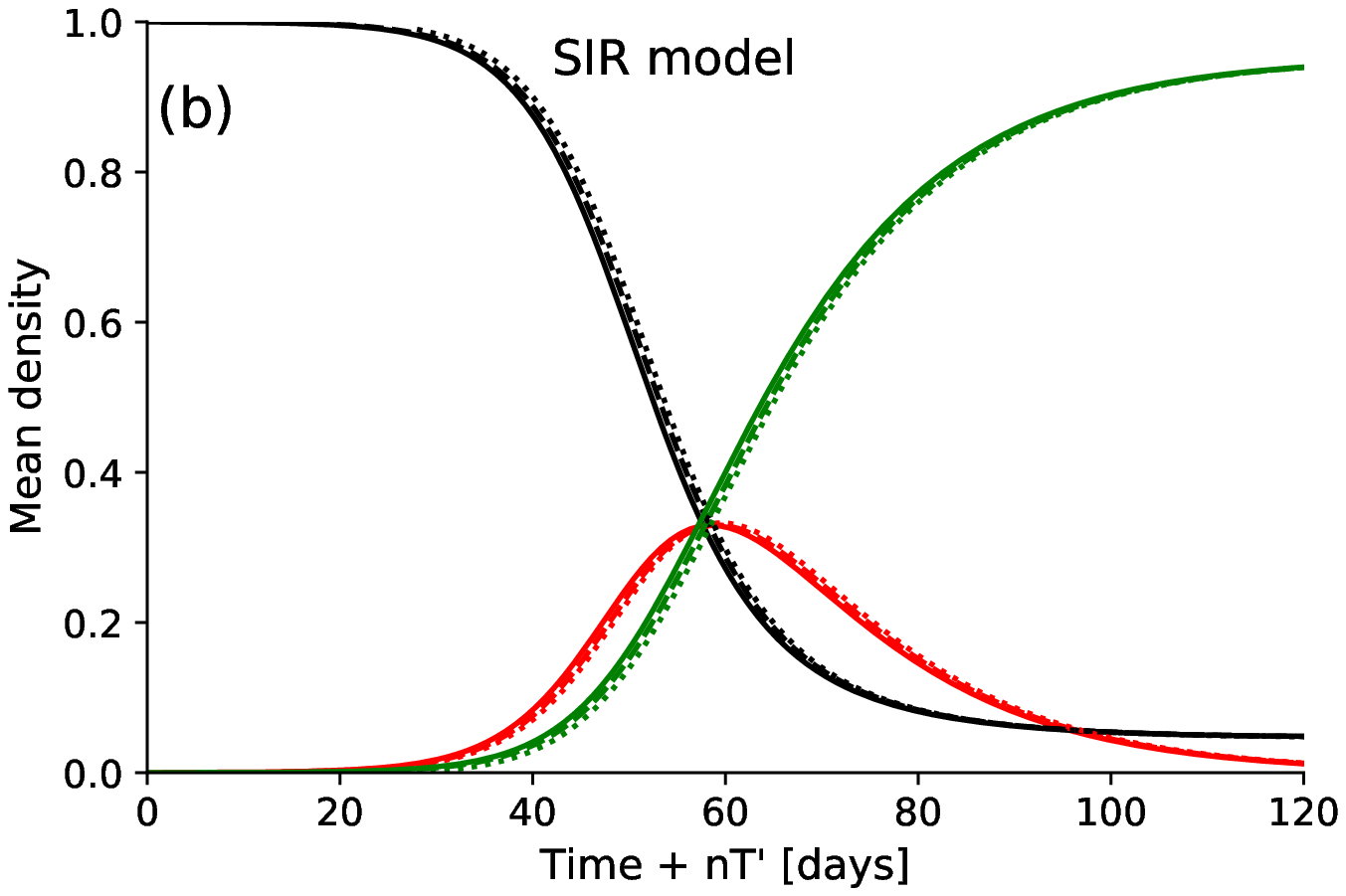}
\end{tabular}

\vskip 0.25cm
\caption{(a) Mean densities of recovered (green lines), active infected (red lines) and susceptible  (black lines) individuals, without epidemic control, obtained from SIR model for 1, 10 and 100 outbreaks of the virus on 100,000 inhabitants (solid, dashed and dotted lines, respectively). (b) collapse for the same data in panel (a) obtained from $nT'\cong n/log(1+\overline{\beta}-1/\overline{t}_{rec})$ and $n=0,\;1,\;2$ for 1, 10 and 100 outbreaks, respectively. For SIR model, all susceptible persons have the same probability of contagion and there is no distinction between them.
} 
\label{FIG3ab}
\end{figure*}

Figure \ref{FIG2} shows the mean densities defined as mean numbers of individuals recovered (green lines), active infected (red lines) and susceptible exposed to the virus (black lines) divided by the total population of 100,000 inhabitants, obtained for 1, 10 and 100 outbreaks of the virus (solid, dashed and dotted lines, respectively). The exposed susceptible individuals are defined as those persons having at least one infected neighbor. For a given number of outbreaks, the mean density curve of these individuals  (black lines) reaches its maximum long before the corresponding infected people peak  (red lines in the same figure). The mean density of susceptible individuals exposed to the virus could be clearly a measure to estimate the probability of contagion.\par

Also, figure \ref{FIG2} shows that infectiousness increases as the number of outbreaks per inhabitant increases and the epidemic peak is earlier. This last result is expected and also predicted by mean field models like Verhulst-Pearle sigmoid \cite{Bizzarri2020} or SIR \cite{Postnikov2020}. Such compartmental models have proven flexible, tractable, and highly informative as a general guide to the
population-level behavior of diseases. Each compartment has either susceptible, infected or recovered persons and the probability of disease-causing contact with any member of a particular compartment is the same. This mean field approximation leads to a fixed intensity of the infection peak i.e., it does not change with the density of outbreaks under 10\% as shown in figure \ref{FIG3ab}(a). Moreover, these results are easily collapsed by a simple translation on the horizontal axis. In figure \ref{FIG3ab}(b), we moved the curves of 10 and 100 outbreaks on the curve of the 1 outbreak, estimating an initial pandemic growth as a geometric progression of common ratio 
\begin{equation}
 r=1+\overline{\beta} - \frac{1}{\overline{t}_{rec}} ,\;\;\mbox{per day.}
\label{ratio}
\end{equation}
Thus, the translation on the horizontal axis is 
\begin{equation}
nT'=n/r , \;\; \mbox{with $n=0,\;1,\; 2$}
\label{translation}
\end{equation}
for 1, 10 and 100 outbreaks, respectively.  The good collapse of the curves is apparent, though a slight difference in the densities of individuals recovered is found at the beginning of their collapse due to every curve starts without individuals recovered (this difference is not visualized in the scale of figure \ref{FIG3ab}(b)). Therefore, for the SIR model, if the density of outbreaks is low enough and known, the pandemic is predictable over time and it is useless for our goal of studying different searching and testing strategies.\par

\begin{figure*}[hbt!]
\centering
\includegraphics[width=8cm]{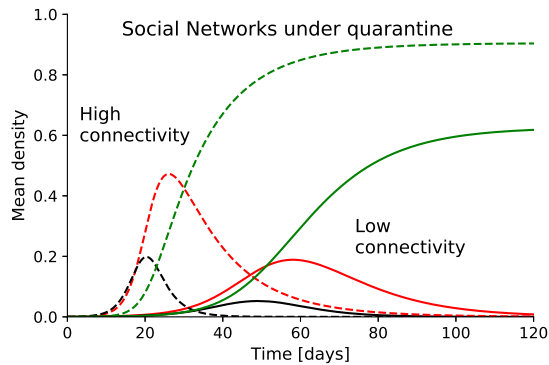} 
\vskip 0.25cm
\caption{Mean densities of recovered (green lines), active infected (red lines) and susceptible exposed to the virus (black lines) individuals, without epidemic control, obtained from 10 outbreaks of the virus on 100,000 inhabitants and social networks under quarantine consisting of $k_{int}=4\pm 2$ cohabitants and two different mean intercommunity connections: $k_{ext}=2$ and $4$ (solid and dashed lines, respectively). The curves were averaged over 100 simulation runs. 
} 
\label{FIG4} 
\end{figure*}

On the other hand, our network model shows a different behavior. The height of the epidemic peak is associated to the network social connectivity. Indeed, a few outbreaks can become extinct without intervention, in areas with few connections and thus, the intensity of the infection peak is low. Epidemic elimination may also be obtained for a higher number of outbreaks when the network connectivity is reduced. This may be clearly observed in figure \ref{FIG4}, where 
the evolution of mean densities of recovered, active infected and susceptible exposed to the virus individuals for two different social structures and a fixed number of outbreaks are shown. The enhancement of connectivity in the network promotes the spread of the disease. Therefore, the social isolation is an effective tool that delays the epidemic peak and also significantly reduces the total number of infected individuals, reflected in the number of recovered individuals (green lines of figure \ref{FIG4}). Since the social isolation has its 
socio-cultural and economic constraints, in the next section, we will apply the algorithm based on the geometry of a hypercube to search and reduce the infection in affected areas of social networks under quarantine.\\

\subsection{Optimal strategies of pool testing to prevent the epidemic spread }
\label{with testing}

The first tests start when a number $n_{I0}$ of infected individuals report to health centres. Each of these persons is considered an infection source and a scanning of their neighbors is done until completing $N/n_{I0}$ individuals around of each infection source, with $N$ the total number of screened persons per day. Once the sample is taken, the testing method is applied. This method has been introduced in \cite{Mutesa2020} and the idea is to pool $N_S$ subsamples of the total sample $N$ and test the combined subsample with a single test. If the test is negative all subjects in the subsample are negative and it continues with another subsample of $N_S$ persons. If the test is positive the hypercube algorithm is applied to determine who are infected.\par
\begin{figure*}[hbt!] 
\centering
\includegraphics[width=8cm]{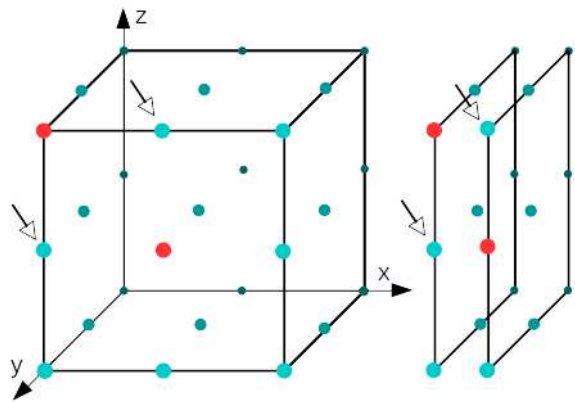}
\caption{Illustration of sample pooling in the hypercube algorithm, for $D = L = 3$ and $N = 27$. Circles in red represent infected persons and the rest in cyan are susceptible. 
Left panel: The hypercube is sliced into $L$ slices, in each of the $D$ principal
directions, and samples from $N/L$ individuals are pooled into a sample for each slice. For this example, 5 slices are positives leading to
4 suspicious persons: the infected individuals and their neighbors pointed with the arrows.
Right panel: the axis $x$ with the maximum number of positive slices is selected. Take one of these, itself
a hypercube of dimension $D - 1$, and run the hypercube algorithm again. The coordinates
of the corresponding infected individual are then uniquely identified, and those of the second
infected individual are inferred by elimination.
}
\label{CUBE}
\end{figure*}

The algorithm consists of locating each individual of the positive subsample on a D-dimensional hypercube lattice with $L$ points in each direction. The hypercube has $D$ principal directions, containing the $N_S$ individuals of the positive subsample, so that $L^D=N_S$. For example, for $D=3$ and $L=3$, the hypercube is a simple cube with $27$ individuals arranged on a $3\times 3\times 3$ grid (figure \ref{CUBE}). Each slice of the hypercube is tested and if is positive the algorithm is running again for the positive slices properly selected to find the infected persons (see \cite{Mutesa2020} for more details of the method). In short, the algorithm is based on the idea that a slice through a D-dimensional hypercube is itself a hypercube of dimension $D-1$.\par 

The effective size $N_S$ of the subgroups is chosen to minimize the total mean number of tests per person. The testing increases as the number of infected individuals increases in the subgroup. Therefore, the algorithm is effective if this number is low and if tests with high sensitivity are used for the dilution of the subsamples, such as reverse-transcription polymerase chain reaction (RT–PCR) tests \cite{Corman2020, Emery2020}. Assuming Poisson statistics for the number of infected individuals in the subgroups and using $L=3$ points in each direction of the hypercube, the optimal size to minimize the total number of tests is \cite{Mutesa2020}
\begin{equation}
 N_S \simeq 0.350/p ,
\label{N_S}
\end{equation}
with $p$ the prevalence defined as the probability that any individual of the subgroup $N_S$ is infected. The prevalence of the disease is unknown and we roughly estimate $p\approx n_{I0}/N$  in the affected areas by the virus, considering that infection does not spread at the beginning of the pandemic. Then, when the testing is finalized, infected individuals in affected areas are found and isolated of their neighbors. Due to limited resources, a maximum number of tests per day $M$ is imposed. $M$ is chosen lower than the sample $N$ of screened individuals per day. The testing is recursively repeated, estimating $p$ as the number of infected individuals isolated, divided by the samples used in the previous testing. In our network model simulations, a fixed number of 15 \% of infected persons report to health centres in each round of testing. The first tests start when the 15 \% of infected individuals is equal or greater than one and thus, $n_{I0}$ depends on the sample.\par

\begin{figure*}[hbt!]  
\centering
\begin{tabular}{cc}
\includegraphics[width=7cm]{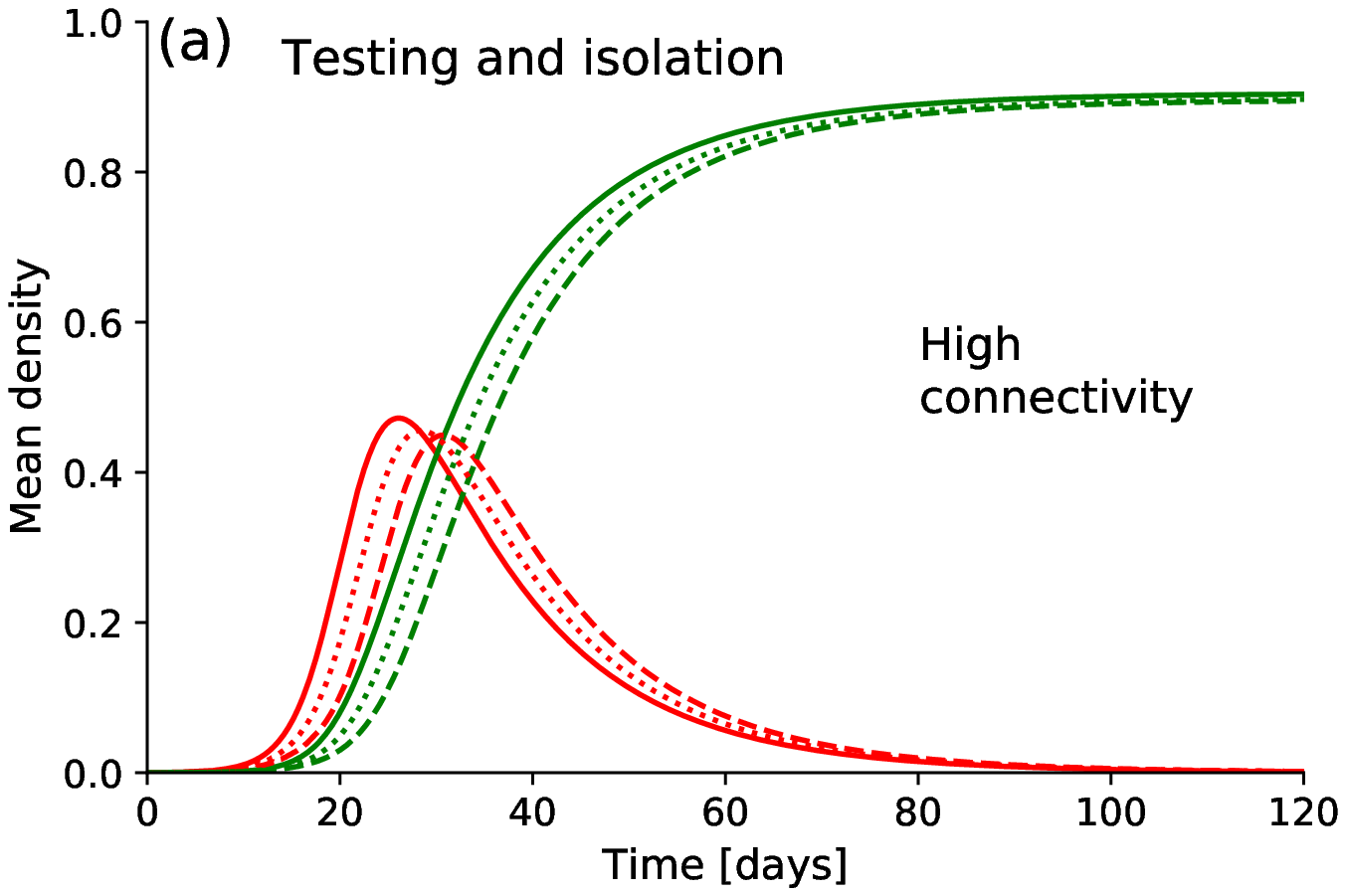} &
\includegraphics[width=7cm]{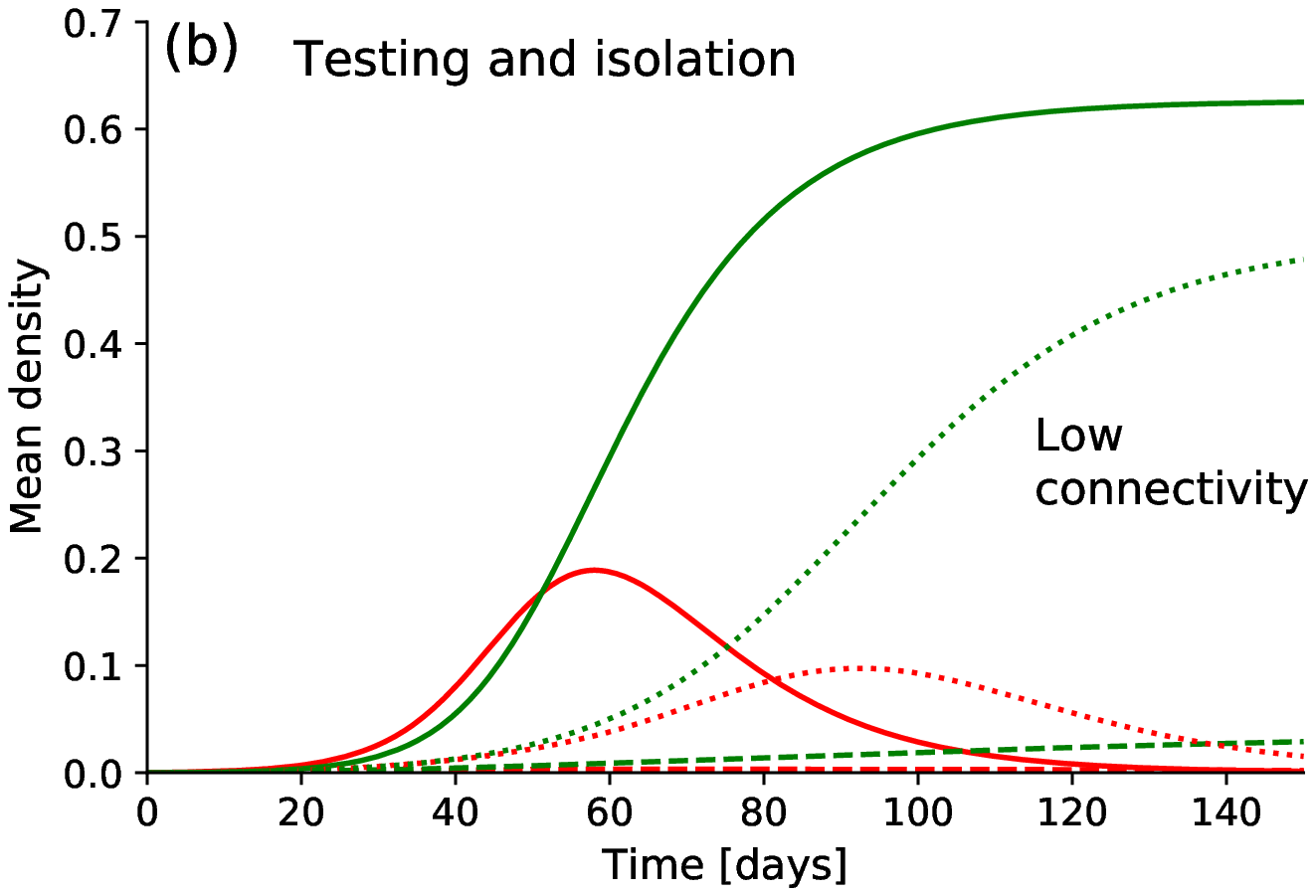}
\end{tabular}   
\includegraphics[width=8cm]{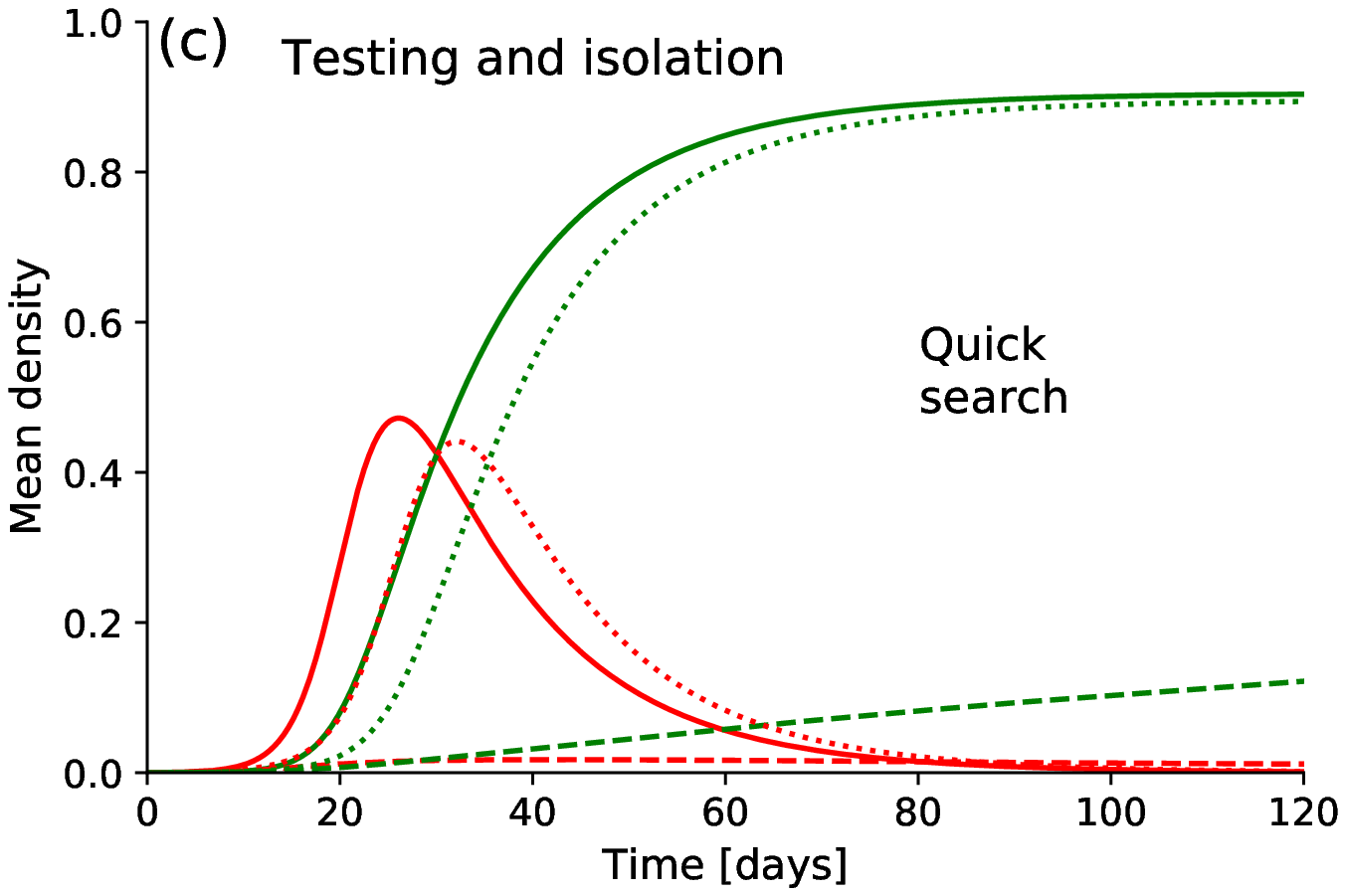}
\vskip 0.25cm
\caption{Comparison of the mean densities of infected (red lines) and recovered (green lines) individuals obtained from social networks under quarantine without testing (solid lines) and with testing and isolation of infected persons using the hypercube algorithm (dashed lines) and the simple method of one test per person (dotted lines) for different scenarios: (a) many connectivity of the networks, $k_{int}=4 \pm 2$ and $k_{ext}=4$, (b) few connectivity of the networks, $k_{int}=4 \pm 2$ and $k_{ext}=2$ and (c) for the same social networks than (a) with a quicker search and isolation of infected persons. The maximum number of tests is $M=200$ on a maximum number of screened individuals $N=2000$ each 100,000 inhabitants per day and in (c) is $N/2$ and $M/2$ each 12 hs to obtain a quicker search, doubling the frequency of testing.   
} 
\label{FIG6}
\end{figure*}

Figure \ref{FIG6}(a) shows the density of active infected and recovered individuals for a social network without epidemic control and with testing using the hypercube algorithm and the simple method of one test per person on affected areas. For both methods of epidemic control,
an epidemic decrease is obtained, significantly improving for social structures with less interaction (figure \ref{FIG6}(b)). In this case, the hypercube algorithm achieves control the epidemic. In fact, the density of recovered persons (dashed green line in figure \ref{FIG6}(b)) is very low since a few individuals were infected (dashed red line, practically is not visualized in the scale of figure \ref{FIG6}(b)).\par

\begin{figure*}  
\centering
\begin{tabular}{cc}
\includegraphics[width=8cm]{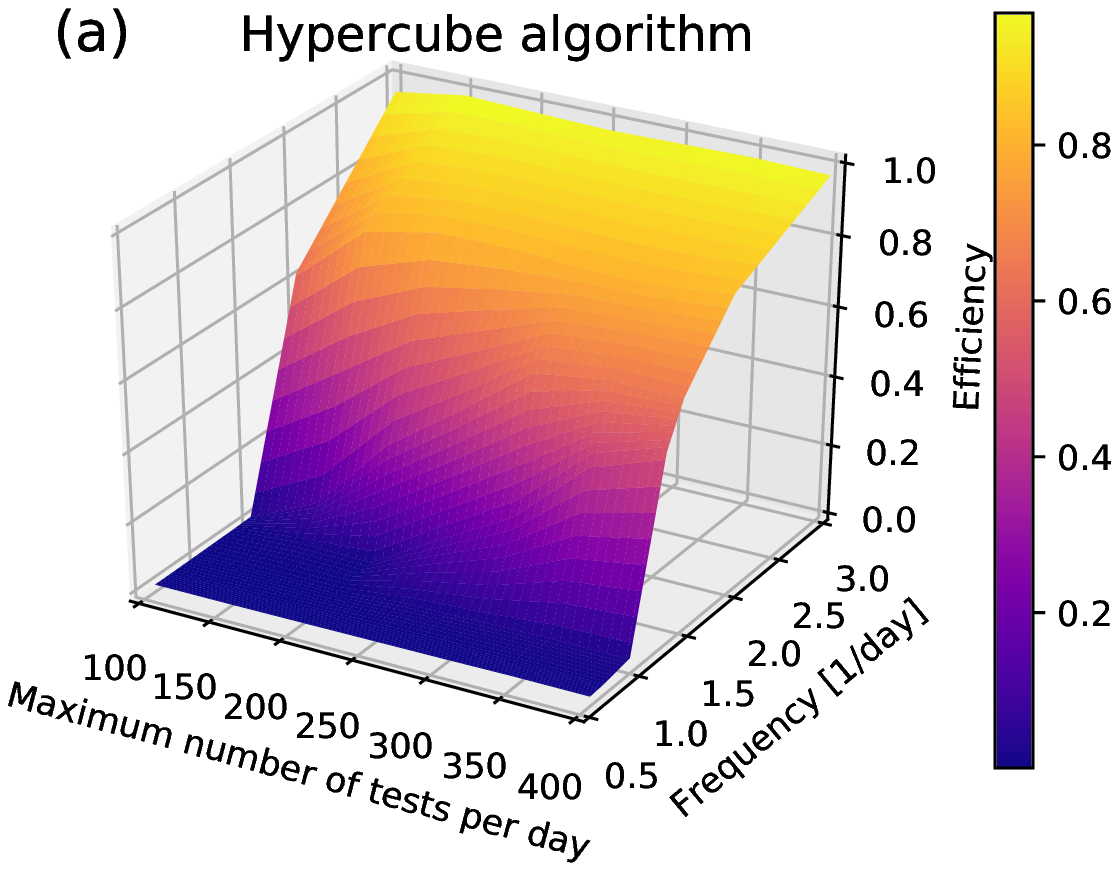}&
\includegraphics[width=8cm]{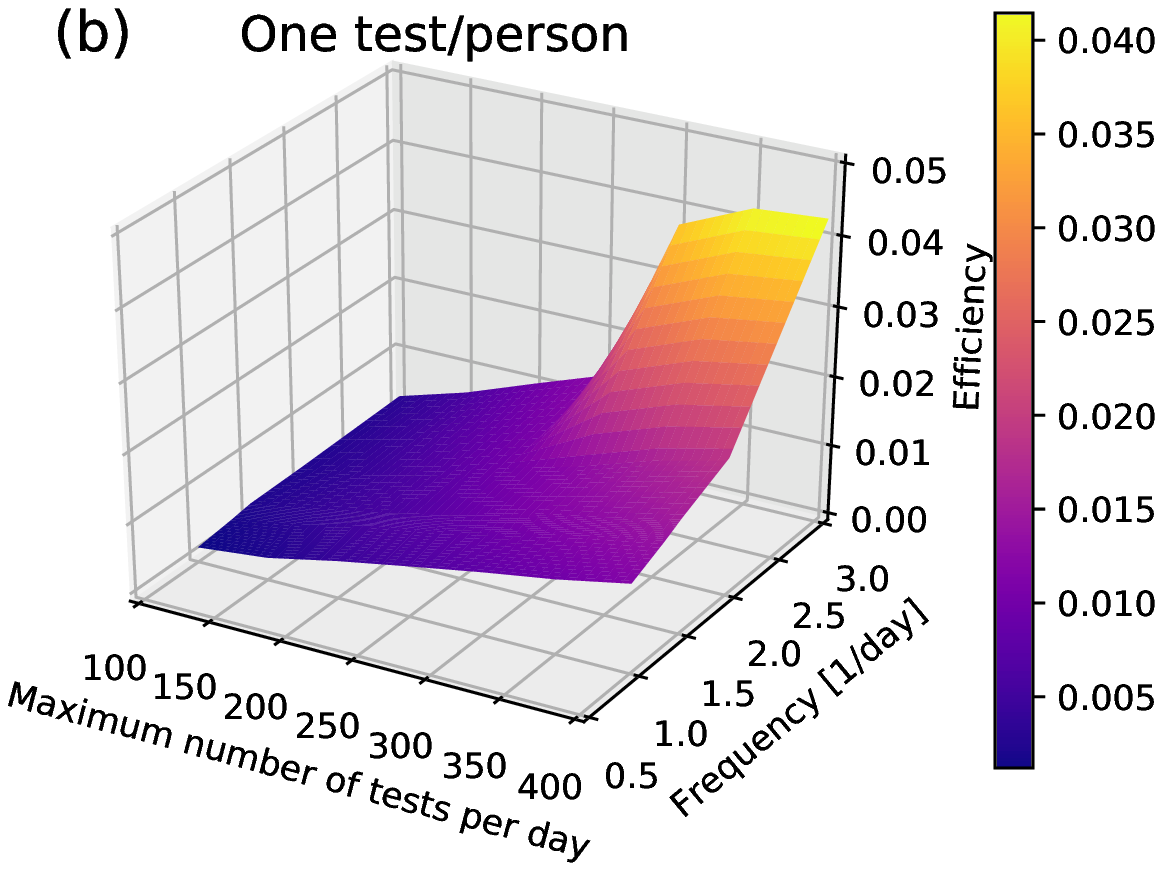}
\end{tabular}
\vskip 0.25cm
\caption{Efficiency of the hypercube algorithm (a) and of the simple method of one test per person (b) as a function of the maximum number of tests per 100,000 inhabitants per day and of frequency of isolation of infected persons for social networks with high connectivity: $k_{int}=4\pm 2$ cohabitants and intercommunity  mean connections $k_{ext}=4$. The hypercube algorithm (a) is much more efficient than one/test person (b). (Note that the efficiency scale in (a) is twenty times higher than in (b)).
} 
\label{FIG7}
\end{figure*}

 Time saving is very important for a rapidly spreading infectious disease like COVID-19. However, the regular testing is limited by costs and operational capacity of sampling. In figures \ref{FIG6} (a)-(c), a maximum number of tests $M=200$ on a maximum number of screened individuals $N=2000$ for every 100,000 inhabitants per day were considered, but in part (c) the maximum number of tests and of screened persons are $M/2$ and $N/2$ every 12 hours, doubling the frequency of testing. This allows the search and isolation of infected individuals in a shorter time and thus, can not infect others. Figure \ref{FIG6} (c) shows that the hypercube algorithm achieves control of epidemic reducing the search time for social networks with high connectivity.\par
 The efficiency of the hypercube algorithm depends on the samples taken from infected individuals report to health centres. If the search of infected individuals is random, the efficiency of the algorithm is low. To check this, we take $M$ persons at random on the network and build hypercubes. Infection curves do not practically change with a random search to both high and low connectivity networks. The search on affected areas is essential to consistently reduce an epidemic like COVID-19. These results are summarized in figure \ref{FIG7}, in which the efficiency of the hypercube algorithm (a) and the simple method of testing every person (b) as a function of the maximum number of tests per day and of frequency of isolation of infected persons is shown, for social networks with high connectivity. The efficiency is defined as the difference between the total number of recovered individuals without epidemic control and with epidemic control divided by the total number of recovered individuals without epidemic control. The maximum number of scanned individuals is ten times the maximum number of tests. The frequency is number of times that infected persons are isolated per day respecting the maximum numbers of tests and scanned individuals (see figure \ref{FIG6} (c)). The efficiency depends on the model of social networks, however the results shown in figure \ref{FIG7}  are qualitatively useful. The frequency of search and isolation of infected individuals on zones reported with virus is the relevant parameter to control the COVID-19. Indeed, a remarkable increment of the algorithm's efficiency is observed when the frequency increases in comparison to an increase of the maximum number of tests per day.

\section{Conclusions}
\label{Conclusions}

The social, psychological and economic burdens throughout a pandemic lead to a minimum threshold of interactions between persons originating a certain confined social structure.
We found that social networks with low connectivity between their individuals reduce the contagion and can go a long way in keeping the curves of infected persons flat. However, since most facets of economic and social life require person-to-person contact, the testing,  searching and isolating infected individuals helps to reduce the epidemics and return sooner to normal activity. RT-PCR (reverse transcription polymerase chain reaction) tests are accurate, but costly and are a challenge particularly for developing countries. The search for infected individuals by grouping samples is considered in this work. Particularly, we studied the epidemic evolution under different strategies of application of a pool testing based on the geometry of a hypercube to isolate infected persons applied to social networks under quarantine threatened by an epidemic with high contagiousness and rapid spread as the coronavirus disease (COVID-19). The pool testing on social networks under quarantine is effective if the search of infected persons is in zones where the virus was reported and the isolation of these individuals is done as quickly as possible. The strategic search in affected zones by the virus and a high frequency of isolation can overcome a massive testing. Indeed, we found that a massive testing randomly applied to social networks with both high and low connectivities leads to little impact on reduction of contagion.


\section{Bibliography}

\end{document}